\documentclass[letterpaper, 10 pt, conference]{ieeeconf}  

\IEEEoverridecommandlockouts                              

\usepackage{graphics}       
\usepackage{epsfig}         
\usepackage{times}          
\usepackage{amsmath}        
\usepackage{amssymb}        
\usepackage{color}
\usepackage{array}

\usepackage{lipsum}
\usepackage{mathtools, cuted}

\title{\LARGE \bf
Dynamics of inertial pair coupled via frictional interface}

\author{Michael Ruderman, Andrei Zagvozdkin, Dmitrii Rachinskii  
\thanks{M Ruderman is with Department of Engineering sciences, University of Agder (UiA).
Postal address: P.B. 422, Kristiansand,4604, Norway. \newline
Email:
        {\tt\small michael.ruderman@uia.no}}
\thanks{A Zagvozdkin and D. Rachinskii are with Department of Mathematical Sciences,
University of Texas at Dallas, Richardson, TX 75080, USA.} }

\begin{document}

\maketitle \thispagestyle{empty} \pagestyle{empty}

\bstctlcite{references:BSTcontrol}

\begin{abstract}
Understanding the dynamics of two inertial bodies coupled via a
friction interface is essential for a wide range of systems and
motion control applications. Coupling terms within the dynamics of
an inertial pair connected via a passive frictional contact are
non-trivial and have long remained understudied in system
communities. This problem is particularly challenging from a point
of view of modeling the interaction forces and motion state
variables. This paper deals with a generalized motion problem in
systems with a free (of additional constraints) friction
interface, assuming the classical Coulomb friction with
discontinuity at the velocity zero crossing. We formulate the
dynamics of motion as the closed-form ordinary differential
equations containing the sign operator for mapping both, the
Coulomb friction and the switching conditions, and discuss the
validity of the model in the generalized force and motion
coordinates. The system has one active degree of freedom (the
driving body) and one passive degree of freedom (the driven body).
We demonstrate the global convergence of trajectories for a free
system with no external excitation forces. Then, an illustrative
case study is presented for a harmonic oscillator with a
frictionally coupled second mass that is not grounded or connected
to a fixed frame. This simplified example illustrates a
realization and main features of the proposed (general) modeling
framework. Some future development and related challenges are
discussed at the end of the paper.
\end{abstract}

\section{Introducing remarks}
\label{sec:1}

Both adhesion/stiction and continuous sliding mechanisms are
essential for modeling a frictional interface between moving
bodies. Dynamic transitions in friction processes, from the
attachment-detachment cycles (see e.g. \cite{zeng2006}) to the
stick-slip and then continuous sliding (or slipping), see e.g.
\cite{socoliuc2004}, have been intensively studied on a physical
level in tribology and material science. Despite available
sophisticated modeling of the frictional interaction of contact
surfaces, from nano- to the meso-scale (see e.g.\
\cite{vanossi2013} and references therein), a pass over to a
lumped parameter modeling of inertial pairs with frictional
interfaces is not trivial. In earlier control and system related
studies, e.g. \cite{armstrong1994}, kinetic friction usually
occurred as an evoked source of damping that acts during the
forced motion. In this case, the causal relationship of the
kinetic friction usually goes from the motion variables as a
source to the friction force variables as a consequence.
Accordingly, an externally excited (in other words actuated)
relative motion is counteracted by generalized frictional forces
arising on a contact interface. This paradigm is entirely
independent of the complexity and level of detail of the friction
modeling. On the other hand, the contact friction can
 occur as a joining interface between two inertial bodies
that are moving and rubbing against each another. Such a causal
relationship would require a generalized frictional force to be
the source, and a relative displacement between two bodies of an
inertial pair in contact to be the result. In this setting, at
least one inertial body is to be regarded as passive and,
therefore, to have an unconstrained frictional interface with
another (driving) body. This interface should allow for both,
conservation of momentum of the moving pair and dissipation of
energy with the associated motion damping. To the best of our
knowledge, such interactions have been less studied in the past.
As such, they deserve attention due to their theoretical and
practical relevance for the motion dynamics and control.

\begin{figure}[!h]
\centering
\includegraphics[width=0.7\columnwidth]{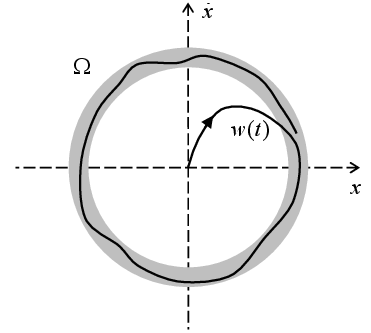}
\caption{A trajectory $w(t)$ of relative motion of the
target body driven via a frictional interface. The ring
$\Omega$ indicates the region of control tolerance within the
$(x,\dot{x})$ phase plane.} \label{fig:1}
\end{figure}

A typical example of an application of inertial systems with a
frictional interface is a periodic motion of the target body put
on a driven surface. Such controlled motion scenarios are
exceedingly common in processing and manufacturing on conveyor
lines and turntables, material flow of items, surface treatment,
preparation of mixed and shaken substances, robot handling of free
(i.e. not flanged) objects, and others. More specifically, a
dynamic motion trajectory $w(t)$ should reach some region of
control tolerance, $\Omega$, surrounding a steady-state periodic
orbit (in the relative $(x,\dot{x})$ coordinates) and stay there
as illustrated in Fig. \ref{fig:1}.
Such control of a periodic motion is sufficiently well understood
for dynamics of rigid bodies and even multi-body systems with
elasticities, despite the fact that non-trivial reference
trajectories and disturbances can pose serious challenges for each
particular application. At the same time, to the best of our
knowledge, a controlled motion along a target trajectory, or (even
simpler) towards a set-point reference, remain largely unexplored
for the manipulation of objects which are solely connected via
frictional interface.

In the following, we focus on the Coulomb friction (on contact
surfaces) \cite{Coulomb1785}. Neither sliding\footnote{Note that
the term `sliding', which is equivalent to `slipping', is
associated here with a continuous macro-motion, i.e. without
changing the sign of relative velocity. These `sliding' regimes or
effects should not be confound with \emph{sliding} dynamics in
Filippov's sense \cite{filippov1988}.} effects of the
Prandtl-Tomlinson type (see e.g. \cite{popov12} for an overview)
nor Stribeck effects \cite{Stribeck1902} will be taken into
account in order to maintain an adequate complexity of modeling
and analysis. We also like to notice that several interesting
historical aspects of the development of the Coulomb dry friction
law and Stribeck friction curves can be found in e.g.
\cite{zhuravlev2013} and \cite{jacobson2003}, respectively.
Furthermore, smooth breakaway frictional force transitions (see
e.g. \cite{Ruderman2017b}) at the start of motion, which are
characteristic for the so-called presliding friction regime (see
e.g. \cite{armstrong1994,AlBender2008}), remain as well outside of
our focus. We should also emphasize that dry Coulomb friction
would become secondary to our analysis if it only acted as a
nonlinear damping and sticking element in an active system with
feedback control (see the recent developments in
\cite{ruderman2021}). It gains main attention here once it appears
as a dynamic interface between two moving bodies. Finally, we
focus on the Coulomb-type frictional interactions in their
simplest form, i.e. with a constant speed-independent magnitude
and the sign opposed to the relative displacement.

The rest of the paper is organized as follows. In Section
\ref{sec:2}, a dynamical framework of an inertial pair connected
via a frictional interface is introduced. Equations
\eqref{eq:2}\,--\,\eqref{eq:4} and Fig. \ref{fig:2} provide the
most general notation of the overall system dynamics, assuming the
contact Coulomb friction with discontinuity at the velocity zero
crossing. The convergence analysis of the free system, i.e.
without exogenous control action, is delivered in Section
\ref{sec:3}, in terms of the system energy (also Lyapunov)
function for the switched and non-switched modes of the relative
motion. We exemplify the problem statement and the proposed
modeling framework by means of an illustrative case in Section
\ref{sec:4}. The discussion and relevant points for our future
developments are summarized in Section \ref{sec:5}.

\section{Dynamics of a pair of bodies with frictional interface}
\label{sec:2}

\begin{figure}[!h]
\centering
\includegraphics[width=0.9\columnwidth]{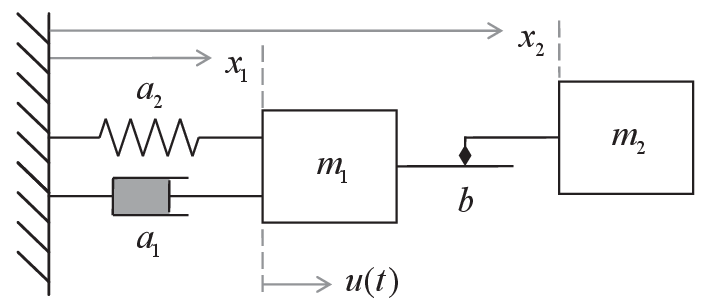}
\caption{A pair of two inertial bodies with a frictional
interface. The first driving inertial body, with the lumped mass
$m_1$, is connected to the ground. The second driven (i.e.
passive) inertial body, with the lumped mass $m_2$, is on a flat
surface. The moving bodies are connected via frictional
interface.} \label{fig:2}
\end{figure}

Let us  consider a pair of
inertial bodies with a frictional interface as shown in Fig.
\ref{fig:2}.
The inertial point-masses $m_1$ and $m_2$ move in the
generalized coordinates $x_1$ and $x_2$, respectively, having
parallel axes. At this point, it does not matter whether a
translational or rotational degree of freedom of the relative
motion is meant. The single restriction is that a relative motion
with only one spatial degree of freedom (DOF) is assumed. A
multidimensional case of, for example, relative motion on a flat
surface, i.e. with two translational and one rotational DOFs, can
be equally elaborated into the proposed modeling framework (see
discussion later in Section \ref{sec:5}), yet it would go
beyond the scope of this paper.

The first (actuated) body is connected to the reference ground by
a virtual spring with the stiffness $a_2$ and a virtual damper
with the viscosity coefficient $a_1$. Note that both virtual
elements can represent (or correspondingly include) mechanical
components as well as feedback control terms. For a feedback
controlled inertial body, $u(t)$ will then constitute an exogenous
input value that can be either a reference motion trajectory, or a
generalized disturbing force, or a combination of both. The second
inertial body is connected to the first body through a frictional
interface which is free of additional constraints. That is, it
experiences unrestricted tangential motion (i.e. no mechanical
limiters as in a backlash case) and no force constraints other
than the normal loading caused by the Coulomb frictional force
$f(t)$. In what follows, we assume the classical Coulomb friction
with the discontinuity at the velocity zero crossing, see e.g. in
\cite{popov2010}. It is worth recalling that the nonlinear Coulomb
frictional force represents a rate-independent damping that can be
understood in combination with infinite stiffness within the
so-called pre-slide friction regime, see \cite{ruderman2017} for
details. While more detailed dynamic friction models attempting to
capture the transient side-effects both during presliding and
sliding have been studied extensively (see e.g. the seminal papers
\cite{armstrong1994, AlBender2008} and the references contained
therein), the basic Coulomb law of friction is often sufficient
for modeling purposes. In this case, the discontinuous Coulomb
friction force opposing the rate of the relative displacement $v$
is captured by $f = b \, \mathrm{sgn}(v)$, where $b$ is the
Coulomb friction coefficient, cf. Fig. \ref{fig:2}. Further we
note that unlike a set-valued sign operator, which is usually used
when dealing with the switching or sliding modes (cf.
\cite{filippov1988}), we use the classical three point valued
signum function of a real number $v$, which is defined as
\begin{equation}\label{eq:1}
    \mathrm{sgn}(v)= \left\{%
\begin{array}{ll}
    1, & \; v>0, \\
    0, & \; v=0,\\
    -1, & \; v<0. \\
\end{array}%
\right.
\end{equation}
This allows well-defined solutions at zero differential velocity
between two moving bodies, which we will use when introducing the
system dynamics.

There are two modes of motion dynamics which should be
distinguished depending on whether the relative displacement
between the inertial bodies occurs or not: (i) when the second
body rests upon the frictional surface (i.e.
$\dot{x}_1=\dot{x}_2$), there is no frictional damping, and the
driving body takes on an additional inertial mass, thus, resulting
in the total $m_1+m_2$; (ii) when the second body slides under the
action of the frictional force, it impacts the momentum of both
bodies which are then either accelerating or decelerating each
other. With these assumptions, the equations of motion of the
coupled pair shown in Fig. \ref{fig:2} can be written as follows:

\newpage
\begin{strip}
\begin{eqnarray}
\label{eq:2}
  x_1-x_2 &=:& z, \\[2mm]
\label{eq:3}
  \Bigl(m_1 + m_2 \bigl(1-\bigl| \mathrm{sgn}(\dot{z}) \bigr| \bigr)  \Bigr) \, \ddot{x}_1 + a_1 \dot{x}_1
  + a_2 x_1 + b\,\mathrm{sgn}(\dot{z}) &=& u(t), \\
\label{eq:4}
   m_2 \ddot{x}_1 \Bigl( 1- \bigl|  \mathrm{sgn}(\dot{z})  \bigr|
   \Bigr) \frac{1}{2} \Bigl(1 - \mathrm{sgn} \bigl( |\ddot{x}_1| - b m_2^{-1} \bigr)
   \Bigr) - m_2 \ddot{x}_2 + b\,\mathrm{sgn}(\dot{z}) &=& 0.
\end{eqnarray}
\end{strip}

It can be seen that an inclusion of the sign operator \eqref{eq:1}
in \eqref{eq:3} and \eqref{eq:4} enables switching directly
between the two above-mentioned modes of the system dynamics. In
both equations, the switching condition is incorporated in the
closed analytic form and, thus, requires neither \emph{if-else}
statements nor \emph{case differences}, that are otherwise usual
when dealing with variable structure hybrid systems. Inclusion of
the switching conditions into the dynamic equations \eqref{eq:3},
\eqref{eq:4} comes, therefore, in favor of the their analysis and
well-defined solutions of the state trajectories. Note that
decoupling of both bodies, correspondingly switching to the mode
(ii), is enabled in \eqref{eq:4}: by the first left-hand-side
bracket for a relative displacement between the bodies, i.e.
$\dot{z} \neq 0$, and by the second left-hand-side bracket for the
case when the stiction condition $|\ddot{x}_1| > b m_2^{-1}$ is
violated.

\section{Convergence analysis of free system}
\label{sec:3}

Let us consider the system \eqref{eq:2}\,--\,\eqref{eq:4} as a
switched system. In the following, we assume that there is neither
viscous damping term nor control, i.e.\ $a_1=0$, $u=0$. The first
assumption is justified by the fact that the viscous damping of
the driving body, i.e. $a_1 \dot{x}_1$, is always dissipative and
does not directly affect the interaction of the inertial pair. The
second assumption of zero control corresponds to  free dynamics of
the system \eqref{eq:2}\,--\,\eqref{eq:4}. Various cases of $u\ne
0$ will be the subject of the future works. With those
assumptions, the system \eqref{eq:2}\,--\,\eqref{eq:4} is
equivalent to
\begin{eqnarray}
\label{e1} \dot x_1&=&v_1, \\
\label{e2} m_1 \dot v_1& =& -a x_1 - b\,\mathrm{sign}(v_1-v_2),\\
\label{e3} m_2 \dot v_2 &= & b\,\mathrm{sign}(v_1-v_2),
\end{eqnarray}
where $v_1=\dot x_1$, $v_2=\dot x_2$ and $a=a_2>0$. The state
space of this system is divided by  the switching surface
$S=\{(x_1,v_1,v_2): v_1=v_2\}$ separating the half-spaces
$\Omega_-=\{(x_1,v_1,v_2): v_1<v_2\}$ and
$\Omega_+=\{(x_1,v_1,v_2): v_1>v_2\}$. The velocity field of
\eqref{e1}\,--\,\eqref{e3} equals
$$\Phi_-(x_1,v_1,v_2)=\left(v_1,-\frac{a}{m_1} x_1+\frac{b}{m_1},-\frac{b}{m_2}\right)\quad \text{in} \quad \Omega_-$$
and
$$\Phi_+(x_1,v_1,v_2)=\left(v_1,-\frac{a}{m_1} x_1-\frac{b}{m_1},\frac{b}{m_2}\right)\quad \text{in} \quad \Omega_+$$
with a discontinuity on $S$. Introducing the energy function
\[
E=\frac{ax_1^2}2+\frac{m_1v_1^2}2+\frac{m_2v_2^2}2,
\]
we observe that
\begin{equation}\label{edot}
\dot E = -b |v_1-v_2|\leqslant 0
\end{equation}
along every trajectory of system \eqref{e1}\,--\,\eqref{e3}.
Hence, the energy is dissipated on those parts of a trajectory
which belong to $\Omega_-\cup \Omega_+$ and is conserved on the
parts which belong to $S$.

\begin{figure}[h]
(a)\includegraphics[width=0.95\columnwidth]{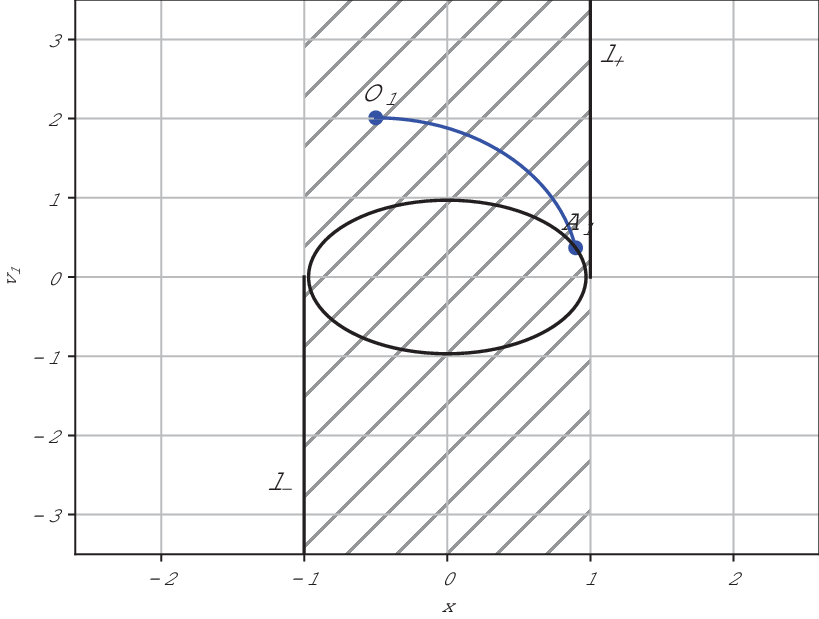}
(b)\includegraphics[width=0.95\columnwidth]{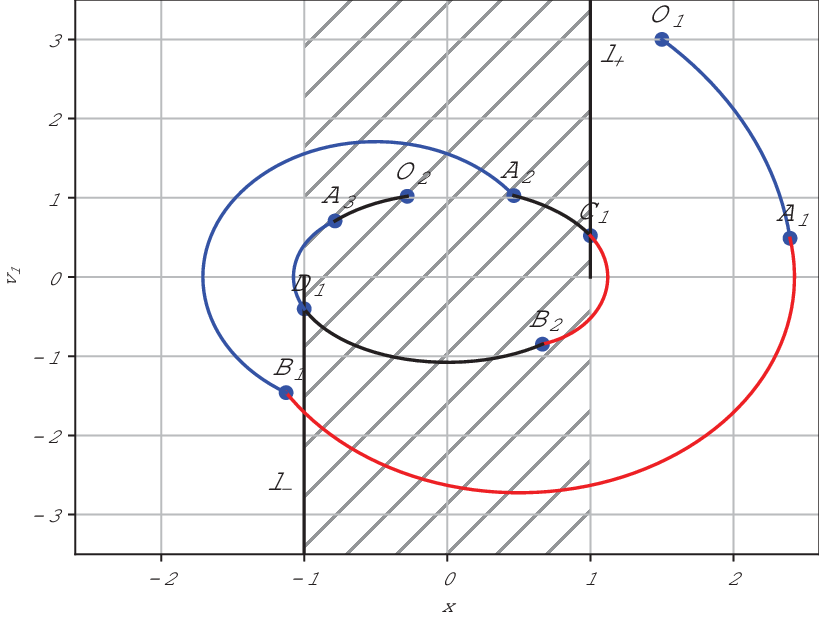}
\centering \caption{Projection of trajectories of system
\eqref{e1}\,--\,\eqref{e3} onto the switching plane $S$. Parts of
the trajectory that belong to the half-spaces $\Omega_+$ and
$\Omega_-$ are shown by blue and red, respectively; parts that
belong to the sliding region $S_0$ of $S$ are shown in black. (a)
A trajectory enters the sliding region $S_0$ (the hatched strip)
of the switching surface $S$ and merges with one of the elliptic
closed trajectories located in $S_0$. The energy of the elliptic
trajectory satisfies $E<E_{cr}$. (b) A trajectory starting at a
point $O_1 \in \Omega_+$ crosses the switching surface $S$ at
point $A_1\in S_+$ and proceeds to the domain $\Omega_-$; crosses
$S$ again returning to $\Omega_+$ at a point $B_1\in S_-$; enters
the sliding region $S_0\subset S$ at a point $A_2$ and proceeds
inside $S_0$ to the exit ray $\ell_+$; exits $S_0$ at a point $C_1
\in l_+$ into the domain $\Omega_-$; re-enters $S$ at a point
$B_2\in S_0$, proceeds inside $S_0$ to a point $D_1\in \ell_-$,
exits $S_0$ through $D_1$ to the domain $\Omega_+$, re-enters
$S_0$ at a point $A_3$ and proceeds inside $S_0$ to a point $O_2$.
When continued, this trajectory converges asymptotically to the
largest elliptic trajectory, which has the energy $E_{cr}$. }
\label{fig:55}
\end{figure}

In order to describe the switching behavior of trajectories, we
consider three parts of $S$:
\[
S=S_-\cup S_0 \cup S_+,
\]
where
\[
S_-=\left\{(x_1,v_1,v_2)\in S: x_1 <- \frac{b (m_1+m_2)}{a
m_2}\right\},
\]
\[
S_0=\left\{(x_1,v_1,v_2)\in S: |x_1| \leqslant \frac{b
(m_1+m_2)}{a m_2}\right\},
\]
\[
S_+=\left\{(x_1,v_1,v_2)\in S: x_1 > \frac{b (m_1+m_2)}{a
m_2}\right\},
\]
and the rays
\[
\ell_-=\left\{(x_1,v_1,v_2): x_1=-\frac{b (m_1+m_2)}{a m_2}, \
v_1=v_2<0\right\},
\]
\[
\ell_+=\left\{(x_1,v_1,v_2): x_1=\frac{b (m_1+m_2)}{a m_2}, \
v_1=v_2>0\right\}
\]
that belong to the boundary of the strip $S_0$. Since the vector
fields $\Phi_\pm$ satisfy
\begin{equation}\label{trans}
\Phi_\pm \cdot n_S<0\ \ \text{on} \ \ S_+, \quad \Phi_\pm \cdot
n_S>0\ \ \text{on} \ \ S_-,
\end{equation}
\begin{equation}\label{trans'}
\Phi_- \cdot n_S\ge 0, \ \ \Phi_+ \cdot n_S\leqslant 0 \ \
\text{on} \ \ S_0,
\end{equation}
where $n_S=(0,1,-1)$ is the normal vector to $S$ pointing from
$\Omega_-$ to $\Omega_+$, the part $S_0$ of $S$ is the
\emph{sliding region} (or surface), see e.g.
\cite{filippov1988,edwards1998,shtessel2014}, consisting of the
trajectories and parts thereof known as \emph{sliding modes}
\footnote{Note that here the term `sliding' is associated with the
sliding modes of dynamics with discontinuities, i.e. in Filippov's
sense \cite{filippov1988}. This should be distinguished from
frictional `sliding', where the inertial bodies (cf.\ Fig.\
\ref{fig:2}) experience a relative displacement with respect to
each other, i.e.\ $\dot{x}_1-\dot{x}_2 \neq 0$. Further, we note
that sliding modes are also well suited for analyzing dynamical
systems with Coulomb friction and feedback controls, see
e.g.~\cite{adly2006finite,ruderman2021}.}. Each of these
trajectories is in the intersection of $S_0$ with an ellipsoid
$E=E_0$ of constant energy. In particular, for
\[
E_0\leqslant E_{cr}=\frac{b^2(m_1+m_2)^2}{2a m_2^2},
\]
the intersection $\{E=E_0\}\cap S_0$ is a closed elliptic
trajectory, see Fig. \ref{fig:55} (a). On the other hand, for
$E_0>E_{cr}$ the intersection $\{E=E_0\}\cap S_0$ consists of two
disjoint elliptic arcs, each being a part of a trajectory. One of
these trajectories exits $S_0$ through the ray $\ell_-$ into
$\Omega_+$, the other through the ray $\ell_+$ into $\Omega_-$.

Since $\dot v_2=-b/m_2<0$ in $\Omega_-$ and $\dot v_2=b/m_2>0$ in
$\Omega_+$, eq. \eqref{edot} implies that
every trajectory 
has infinitely many intersections with the switching plane $S$.
From \eqref{trans} it follows that any trajectory from $\Omega_-$
either enters the sliding region $S_0$ of $S$ and proceeds as
described above or intersects $S$ at a point $p\in S_-$
transversally (i.e.\ the intersection point is isolated) and
proceeds to $\Omega_+$. Similarly, any trajectory from $\Omega_+$
either enters the sliding region $S_0$ or proceeds to $\Omega_-$
transversally through the part $S_+$ of $S$, see Fig. \ref{fig:55}
(b).

Due to these considerations, eqs. \eqref{edot}, \eqref{trans}
imply that any trajectory of system \eqref{e1}\,--\,\eqref{e3}
either merges with one of the elliptic periodic trajectories
(sliding modes) $\{E=E_0\}\cap S_0$ with $E_0\leqslant E_{cr}$ in
finite time or converges to the largest of the elliptic
trajectories, $\{E=E_{cr}\}\cap S_0$, asymptotically. In the
latter scenario, there is a sequence of times
$t_1<\tau_1<t_2<\tau_2<\cdots$ such that
\[
\tau_k-t_k\to \pi \sqrt{\frac{m_1+m_2}{a}}, \quad
t_{k+1}-\tau_k\to 0
\]
as $k\to\infty$ and the trajectory belongs to the sliding region
$S_0$ during each time interval $[t_k,\tau_k]$, i.e.\ the
trajectory leaves $S_0$ only for short intervals of time; further,
all the exit points from $S_0$ and entry points to $S_0$ are located
near the end points $(\pm b (m_1+m_2)/(a m_2), 0,0)$ of the rays
$\ell_\pm$. Also, equation \eqref{edot} implies that $x_1-x_2\to
const$ for any trajectory.

We conclude that the set of periodic trajectories $\{E=E_0\}\cap
S_0$ with $E_0\leqslant E_{cr}$ (including the equilibrium at the
origin) is globally asymptotically stable. In particular, one can
show that any trajectory starting from $S_0$ with energy $E_0$,
which is slightly higher than $E_{cr}$, converges to the periodic
trajectory $\{E=E_{cr}\}\cap S_0$ without merging with it. This
convergence is slow, slower than exponential. In particular, the
energy of such solutions satisfies $c_1(E(0))/t \le E(t)-E_{cr}\le
c_2(E(0))/t$ with $0<c_1(E(0))<c_2(E(0))$.

It is further worth noting, that when a relatively small viscous
friction is present, the switching dynamics is similar but the
trajectories inside the sliding region spiral towards the
equilibrium at the origin, which is globally stable in this case.
The counterpart of eq. \eqref{edot} reads $\dot E = -b
|v_1-v_2|-a_1 v_1^2$.

\section{Illustrative case study}
\label{sec:4}

\subsection{Example system}
\label{sec:4:0}

The proposed dynamic framework \eqref{eq:2}\,--\,\eqref{eq:4} is
further elucidated by the following simplified case study. We
assume both unit masses, and the coefficients of the linear
sub-dynamics from \eqref{eq:3} are assumed to be $a_1 = 0$ and
$a_2 = 200$. This renders the first moving body to be a harmonic
oscillator, which has the eigenfrequency of $10$ rad/s when the
second body is not sliding and rests upon the first one. We allow
for different values of the Coulomb friction coefficient to be $0
< b < 1$. Assuming a free motion, i.e.\ $u(t)=0$, the dynamics
\eqref{eq:3}\,--\,\eqref{eq:4} reduces to an autonomous system
\begin{eqnarray}
\label{eq:5}
\ddot{x}_1 & = & - \Bigl( 2- \bigl| \mathrm{sgn}(\dot{z}) \bigr| \Bigr)^{-1}  \Bigl(200 \, x_1 + b \,\mathrm{sgn}(\dot{z})\Bigr) , \\
\label{eq:6} \ddot{x}_2 & = & \Bigl( 1 - \bigl|
\mathrm{sgn}(\dot{z}) \bigr| \Bigr) \, \ddot{x}_1 + b \,
\mathrm{sgn}(\dot{z}).
\end{eqnarray}
Note that \eqref{eq:6} does not include the acceleration-dependent
switching since the system ensures that $|\ddot{x}_1| < b
m_{2}^{-1}$, cf.\ \eqref{eq:4}. An initial condition $x_1 \, \vee
\, \dot{x}_1 \neq 0$ ensures a finite energy storage at $t=t_0=0$
and, therefore, the onset of the relative motion of the free
oscillator. According to \eqref{eq:6}, the second mass is either
synchronized with the first inertial mass and, then, has the same
acceleration when sticking, or the second mass is sliding with
respect to the first mass and, thus, dissipating energy when
$\mathrm{sgn}(\dot{z}) \neq 0$. In order to develop the state
trajectory solutions, we need to distinguish between these two
modes of the relative motion:
\begin{eqnarray}
\label{eq:7}
  \left.%
\begin{array}{ll}
    2 \, \ddot{x}_1 + 200 \, x_1 = 0 \\
    \ddot{x}_2 =  \ddot{x}_1 \\
\end{array}%
\right \} &  \hbox{if } \; \mathrm{sgn}(\dot{z}) = 0, \\[2mm]
\label{eq:8}
  \left.%
\begin{array}{ll}
    \ddot{x}_1 + 200 \, x_1 = \mp b \\
    \ddot{x}_2 = \pm b \\
\end{array}%
\right \} &  \hbox{if } \; \mathrm{sgn}(\dot{z}) \neq 0.
\end{eqnarray}
When inspecting the switched system dynamics \eqref{eq:7},
\eqref{eq:8}, one can see that in the first mode, i.e.\
\eqref{eq:7}, the system behaves as a pair of synchronized
undamped harmonic oscillators with $x_1-x_2=\mathrm{const}$ and
$\dot x_1=\dot x_2$ (i.e.\ $\dot z=0$). On the other hand, the
piecewise linear dynamics \eqref{eq:8} leads to the damped
oscillations of $(x_1,\dot{x}_1)$ and, thus, to a synchronization
of the orbits of both inertial bodies, i.e. $x_1(t)-x_2(t)
\rightarrow \mathrm{const}$ and $\dot{z} \rightarrow 0$, which is
in line with the general results obtained in Section \ref{sec:3}.

\subsection{Numerical results}
\label{sec:4:1}

The following numerical results were obtained for system
\eqref{eq:5}, \eqref{eq:6} with the first-order forward Euler
solver.

\begin{figure}[!h]
\centering
(a)\includegraphics[width=0.95\columnwidth]{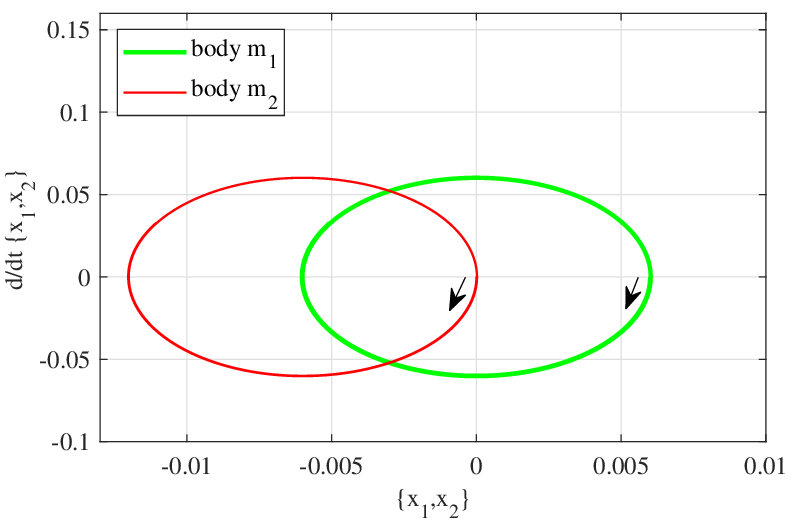}
(b)\includegraphics[width=0.95\columnwidth]{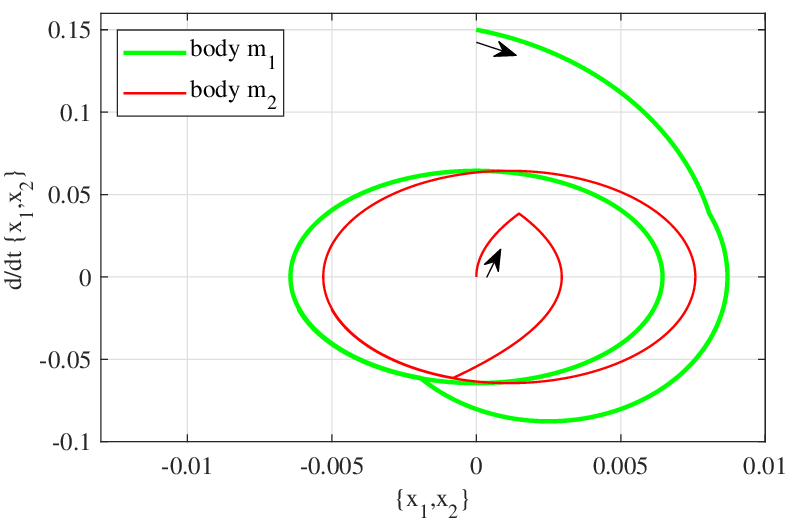}
\caption{The motion trajectories (in the $(x,\dot{x})$
phase-plane) of the system \eqref{eq:5}, \eqref{eq:6}: with the
initial $x_1(t_0)=0.006$ in (a), and the initial
$\dot{x}_1(t_0)=0.15$ in (b). Other initial values are set equal
to zero; the Coulomb friction coefficient is $b=0.5$.}
\label{fig:3}
\end{figure}

Figure \ref{fig:3} shows the motion trajectories of both inertial
bodies in the $(x,\dot{x})$ phase-plane when the Coulomb friction
coefficient is set to $b=0.5$. Motion trajectories with an initial
value $x_1(t_0)=0.006$ are depicted in the plot (a); the other
initial values are zero. One can see that the system \eqref{eq:5},
\eqref{eq:6} starts in the stiction mode \eqref{eq:7} and remains
in this mode at all times featuring undamped harmonic oscillations
with the synchronized orbits $(x_1,\dot{x}_1)$ and
$(x_2,\dot{x}_2)$. On the other hand, when an initial value
$\dot{x}_1(t_0)=0.15$ is used, as shown in the plot (b), the
system \eqref{eq:5}, \eqref{eq:6} starts in the mode \eqref{eq:8},
i.e. the bodies are in a relative motion subject to the frictional
damping. Further, after a period of time the motion of both bodies
synchronizes as $\dot{z}$ converges to zero. This transition to
synchronization is illustrated by the next example.

\begin{figure}[!h]
\centering
(a)\includegraphics[width=0.95\columnwidth]{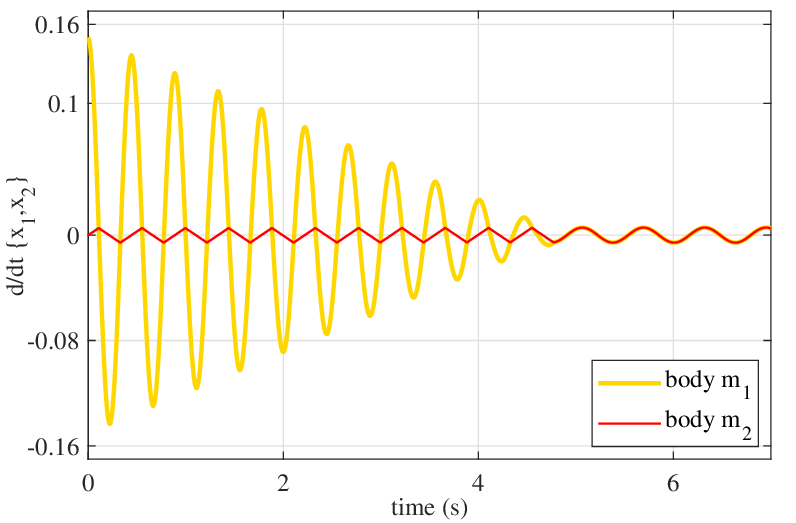}
(b)\includegraphics[width=0.95\columnwidth]{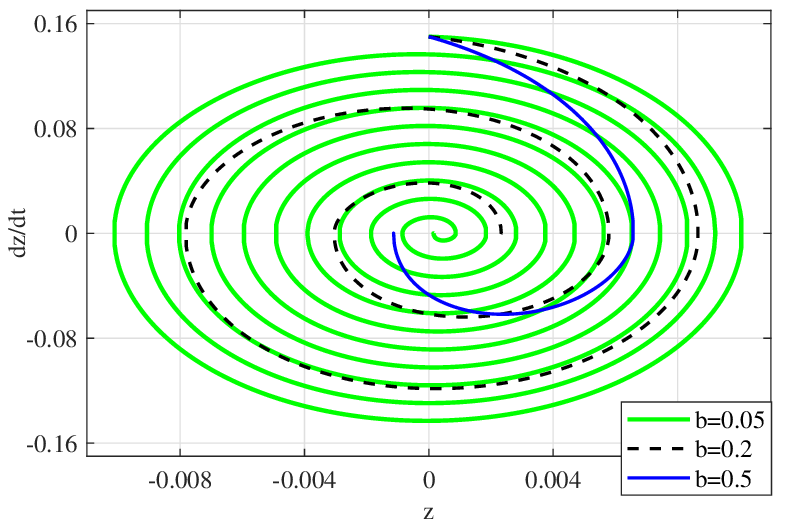}
\caption{Motion trajectories of system \eqref{eq:5},
\eqref{eq:6} with the initial value $\dot{x}_1(t_0)=0.15$;
the other initial values are zero; the time series $\dot{x}_1(t)$ versus $\dot{x}_2(t)$
for $b=0.05$ in (a), and the $(z,\dot{z})$ phase portrait for
$b=\{0.05,0.2,0.5\}$ in (b).} \label{fig:4}
\end{figure}

Simulations of the trajectories with an initial value
$\dot{x}_1(t_0)=0.15$ for different values of the Coulomb friction
coefficient $b=\{0.05,0.2,0.5\}$ are shown in Fig. \ref{fig:4}
(b); the panel (a) exemplifies the corresponding time series of
$\dot{x}_1(t)$ and $\dot{x}_2(t)$ for $b=0.05$. One can recognize
a constant damping rate of $\dot{x}_1(t)$ oscillations due to the
Coulomb type energy dissipation. The $\dot{x}_2(t)$ trajectory
proceeds as a saw-shaped oscillation, owing to the constant
acceleration $\ddot{x}_2=\pm b$ until both bodies synchronize as
$\dot{z}$ approaches zero at the time $t \approx 4.7$ sec. The
convergence of the $(z,\dot{z})$ state towards the invariant set
$\dot z=0$ is clearly visible in Fig. \ref{fig:4} (b) for the
three different values of the Coulomb friction coefficient.

\section{Discussion}
\label{sec:5}

The modeling framework, analysis, and examples described in this
paper provide a transparent systems- and control-oriented view of
how the friction interface of an inertial pair realizes the
coupling of interacting forces and motion variables. One simple
insight is that the dry Coulomb friction law can be directly used
for defining the switched system dynamics with the two modes of
relative motion: (i) the bodies stick to each other; (ii) the
bodies slide against each other, cf.
\eqref{eq:2}\,--\,\eqref{eq:4} and \eqref{eq:7}, \eqref{eq:8}.
Moreover, the fact that the moved passive body (cf. Fig.
\ref{fig:2}) has an unbounded motion space gives an additional
insight into how such an underactuated motion system could be
controlled when the active body appears as a single (external)
energy source, while the motion variables of interest are $(x_2,
\dot{x}_2)$. A few more discussion points are in order.

\begin{itemize}
    \item The free system \eqref{eq:2}\,--\,\eqref{eq:4}
    with $u=0$ is subject to synchronization in the relative $(x_1,
    \dot{x}_1)$ and $(x_2, \dot{x}_2)$ coordinates. In particular,
    for any initial conditions, the global asymptotic convergence
    $\dot{z}(t) \rightarrow 0$ for $t \rightarrow \infty$
    is warranted for all the admissible (i.e. physical) system parameters
    $m_1, m_2, a_2, b > 0$ and $a_1 \geq 0$.

    \item Zero convergence of $\dot{z}(t)$ and, therefore, a full
    synchronization of both moving bodies does not take place in a
    single dynamic mode, cf. Section \ref{sec:3} and Fig.
    \ref{fig:4} (a). A finite time convergence towards a synchronized
    orbit can precede the following long-term alternation
    of the stiction (i.e. adhesion) and slipping modes, both in the relative
    $(z,\dot{z})$ coordinates.

    \item The viscous damping of the active (driving) body influences
    the trajectories of the overall system, but has no principal
    impact on the synchronization mechanism of the moving bodies in a
    friction-coupled inertial pair. At the same time, a purposefully
    controlled viscous damping of the active body (cf. damping with $a_1$ in Fig.
    \ref{fig:2}) can accelerate the synchronization and be
    purposefully used for  trajectory tracking, cf. Fig. \ref{fig:1}.

    \item We used two alternative descriptions
    of motion of the coupled inertial pair. As we have seen, the switched system
    \eqref{eq:2}\,--\,\eqref{eq:4} allows for zero velocity
    solutions in $\dot{z}$, cf. \eqref{eq:1}, i.e. adhesion of both
    moving bodies to each other. On the other hand, the
    global asymptotic convergence of the free system was shown (cf. Section
    \ref{sec:3}) by considering the switched system
    in Filippov's sense  (cf.\ \eqref{e1}\,--\,\eqref{e3}),
    thus allowing for the sliding modes.
    In the authors' opinion, both interpretations of the system switching are possible
    for the Coulomb friction with a discontinuity. In particular,
    the sign operator as defined in \eqref{eq:1} interprets
    the adhesion of inertial bodies to each other in a proper
    physical sense of zero relative velocity. As such, it enables a closed
    analytic form of the modeling framework
    \eqref{eq:2}\,--\,\eqref{eq:4}.

    \item While we considered systems with one spatial
    degree of freedom (in the generalized coordinates), the proposed modeling
    framework can be further extended to describe motions on a surface.
    In such a case, one needs to use both
    orthogonal translational coordinates, e.g. $(x,y)$,
    and rotational coordinates, e.g. $\varphi$, and introduce
    an appropriate vector of Coulomb
    friction coefficients, $\mathbf{b}_{x,y,\varphi}$.
    Furthermore, this setting requires careful modeling because
    the coupling terms of friction on a surface
    are non-trivial from the tribological viewpoint.

    \item The switched motion system
    \eqref{eq:2}\,--\,\eqref{eq:4} is underactuated as well as
    upped-bounded by $\ddot{x}_1 = \ddot{x}_2$, as for the control efforts, cf.
    \eqref{eq:7}. The energy- and performance-efficient control methods for the class of
    friction-coupled dynamical systems
    \eqref{eq:2}\,--\,\eqref{eq:4} are expected to be challenging.
    A case study of system \eqref{eq:2}\,--\,\eqref{eq:4}
    with different types of the application-motivated control, in terms of the dynamic
    reference signals and external disturbances, will be the subject of the future works.
    Also, the effect of the time- and state-varying
    behavior of the frictional interface, i.e. $b(\cdot)$,
    on the motion dynamics poses an interesting open problem.
\end{itemize}

\bibliographystyle{IEEEtran}
\bibliography{references}

\end{document}